\documentclass[12pt]{article}
\usepackage{setspace}
\usepackage[round,numbers]{natbib}
\usepackage{arxiv}
\usepackage{graphicx}
\usepackage{bm}
\usepackage{amsmath}
\usepackage{amssymb}
\usepackage{comment}
\usepackage{bbm}
\usepackage{float}
\usepackage{subcaption}
\usepackage{chngpage}
\usepackage{gensymb}
\usepackage{url}
\usepackage{authblk}
\usepackage{blindtext}
\makeatletter
\renewcommand\@biblabel[1]{#1.}
\makeatother

\usepackage{mathtools}

\newtagform{brackets}{[}{]}
\usetagform{brackets}

\newcommand{\given}{\, | \,}

\newcommand{\calE}{{\cal E}}

\newcommand{\calG}{{\cal G}}
\newcommand{\calV}{{\cal V}}

\begin{document}
\date{}
\title{Spatial Modeling for Correlated Cancers Using Bivariate Directed Graphs}

\author{Leiwen Gao\thanks{Department of Biostatistics, University of California Los Angeles \  gaoleiwen@ucla.edu}}
\author{Sudipto Banerjee\thanks{University of California Los Angeles \ sudipto@ucla.edu}}
\author{Abhirup Datta\thanks{Johns Hopkins University \ abhidatta@jhu.edu}}
\affil

\maketitle
\noindent \textit{Contributions:} (I) Conception and design: All authors; (II) Administrative support: S Banerjee; (III) Provision of study materials or patients: Not applicable (IV) Collection and assembly of data: L Gao; (V) Data analysis and interpretation: L Gao; (VI) Manuscript writing: All authors; (VII) Final approval of manuscript: All authors

\begin{abstract}
\textbf{Background:} Disease maps are an important tool in cancer epidemiology used for the analysis of geographical variations in disease rates and the investigation of environmental risk factors underlying spatial patterns. Cancer maps help epidemiologists highlight geographic areas with high and low prevalence, incidence, or mortality rates of cancers, and the variability of such rates over a spatial domain. They can also be used to detect ``hot-spots'' or spatial clusters which may arise due to common environmental, demographic, or cultural effects shared by neighboring regions. Statistical methods for spatial data formulate models to capture spatial autocorrelation and produce cancer maps to better detect clustering and hotspots. When more than one cancer is of interest, the models must also capture the inherent or endemic association between the diseases in addition to the spatial association. This article develops interpretable and easily implementable spatial autocorrelation models for two or more cancers.  

\textbf{Methods:} The article builds upon recent developments in univariate disease mapping that have shown the use of mathematical structures such as directed acyclic graphs to capture spatial association for a single cancer. The advantage of using directed acyclic graphs over other existing models is the easier interpretation of spatial association. The current manuscript extends this family of directed acyclic graphical models to estimate inherent or endemic association for two cancers in addition to the association over space (clustering) for each of the cancers. The method builds a Bayesian hierarchical model where the spatial effects are introduced as latent random effects for each cancer. A valid joint probability model is constructed by first modeling the marginal distribution of one disease followed by the second disease conditional on the first. This approach ensures easier interpretation of model parameters and helps to separate the spatial autocorrelation for each cancer from the association between the two cancers.   

\textbf{Results:} We analyze the relationship between esophagus and lung cancer extracted from the Surveillance, Epidemiology, and End Results (SEER) Program for their incidence rates in the years 2012 - 2016 across 58 counties in California. Our analysis shows statistically significant association between the county-wise incidence rates of lung and esophagus cancer across California. After accounting for the explanatory variables, esophagus cancer rates exhibit weaker spatial association than lung cancer rates for data counties in California.

\textbf{Conclusions:} The bivariate directed acyclic graphical model performs better than competing bivariate spatial models in the existing literature. This improvement is seen both in terms of the model's fit to the data and complexity of the model.
\end{abstract}	

	
\section{Introduction}

\noindent \noindent %\textbf{\emph{Importance and critical barriers:}} 
Disease mapping, which refers to techniques for mapping and analysis of geographical variations in disease rates and the investigation of environmental risk factors underlying these patterns, has long been an important tool in cancer epidemiology \citep{koch2005cartographies}. Disease maps are used to highlight geographic areas with high and low prevalence, incidence, or mortality rates of cancers, and the variability of such rates over a spatial domain. They can also be used to detect ``hot-spots'' or spatial clusters which may arise due to common environmental, demographic, or cultural effects shared by neighboring regions. Maps of crude incidence or mortality rates can be misleading when the population sizes for some of the units are small, which results in large variability in the estimated rates, and makes it difficult to distinguish chance variability from genuine differences. The correct geographic allocation of health care resources can be greatly enhanced by deployment of statistical models that allow a more accurate depiction of true disease rates and their relation to explanatory variables (covariates). Many tasks critical for successful cancer surveillance and control require new inferential methods to handle these complex and often spatially indexed data sets. Since local sample sizes within each spatial region are too low for design-based solutions to attain desired levels of statistical precision \citep{schaible2013indirect}, much recent work in disease-mapping has been carried out within the context of Bayesian hierarchical models \citep{banerjee2014hierarchical}. The body of scientific literature on modern methods for geographic disease mapping is too vast to be reviewed here. Comprehensive reviews of prevalent statistical disease mapping methods and their implementation using available software can be found, among several other sources \citep{best2005comparison,waller2010handbook,waller2004applied,lawson2013statistical}.  

Statistical models for mapping a single disease have employed probability distributions such as Markov random fields or MRFs \citep{rueheld04} that introduce dependence using the adjacency information among the different regions on a map. Two popular examples are the Conditional Autoregression (CAR) and Simultaneous Autoregression (SAR) models \citep{besag1974spatial, besag1991bayesian, anselin1988lagrange, kissling2008spatial} for further discussions on CAR and SAR models. More recently, Directed Acyclic Graphical Autoregressive (DAGAR) models that employ directed acyclic graphs have been developed as a preferred alternative to CAR or SAR models \citep{datta2018spatial}. A specific motivation for DAGAR models is that they impart greater interpretability to the spatial autocorrelation parameter.

In this article, we will perform joint spatial mapping of two different types of cancers. Joint modeling is appropriate when different diseases have been observed over the same spatial units and when the diseases themselves are related to each other, say because they share the same set of spatially distributed risk factors or the presence of one disease in a spatial unit may encourage or inhibit the presence of the second disease in the same spatial unit. Put differently, we seek models to capture the spatial association for each disease as well as the association between the diseases. There is, by now, a substantial literature on multivariate disease mapping \citep{knorr2001shared, kim2001bivariate, gelfand2003proper, carlin2003hierarchical, held2005towards, jin2005generalized, jin2007order,diva2008parametric, zhang2009smoothed, martinez2013general,mari2014smoothed}. These articles have demonstrated, theoretically and empirically, the benefits of jointly modeling several potentially related cancers, as opposed to modeling them independently. While it has been assertively demonstrated that independent models for cancers can lead to biased results because of unaccounted associations among the cancers, the current literature is largely based on using CAR models for spatial mapping.  Our proposed bivariate DAGAR model for modeling two diseases over the same spatial region will help epidemiologists and spatial analysts better interpret the association among the cancers. 

The balance of this article proceeds by developing a class of bivariate DAGAR models, conducting some disease mapping for two different cancers, and summarizing  with some concluding remarks.

\section{Methods}\label{sec: DAGAR}
Our approach will be to construct a probability model for each disease using the distribution specified by DAGAR. We will extend the univariate DAGAR to a bivariate model by modeling the distribution of one disease as a univariate DAGAR and the conditional distribution of the second disease given the first also as a DAGAR. In this sense, our bivariate DAGAR is analogous to the bivariate CAR models \citep{jin2005generalized}. We develop notations and briefly discuss the univariate DAGAR in the next section, followed by the bivariate extension in the following section. 

\subsection{DAGAR for modeling a single disease}\label{sec: univariate} 
We consider a geographic map of our region of interest (e.g., a particular state) delineated by $k$ distinct administrative regions (e.g., counties or ZIP codes) with clear boundaries separating them. Let $w = (w_1, w_2, \dots, w_k)^\top$ be a $k\times 1$ vector consisting of spatially associated random effects corresponding to each region. We develop a spatially correlated model using a directed acyclic graph. The geographic map provides us with a list of neighbors for each region. Neighbors can be defined by the user. Common definitions include when two regions share a common boundary or if their centers are within a certain fixed distance, although the model and resulting distribution theory hold for any fixed set of neighbors. The data structure for the geographic map and its neighbors is defined as a \emph{graph}, denoted $\calG = \{\calV, \calE\}$, where the regions are indexed by an ordered set $\calV = \{1,2,\ldots,k\}$ and form the vertices of the graph and $\calE$ is the collection of edges between the vertices, i.e., the collection of ordered pairs $(j,j')$ such that $j$ and $j'$ are geographic neighbors based upon some specified definition. 

The DAGAR model specifies $w \sim N(0, \tau Q(\rho))$, where $Q(\rho)$ is a spatial precision matrix that depends only upon a spatial autocorrelation parameter $\rho$ and $\tau$ is a positive scale parameter. To describe $Q(\rho)$, we define neighbor sets $N(i) = \{j < i: j \sim i\}$, where $i \in \calV \setminus \{1\}$, i.e. the set $\calV$ excluding the region indexed by $1$, and $j\in \calV$. Thus, $N(i)$ includes geographic neighbors of region $j$ that \emph{precede} $i$ in the ordered set $\calV$. The precision matrix $Q(\rho) = (I-B)^{\top}F(I-B)$, where $B$ is a $k\times k$ strictly lower-triangular matrix with entries $b_{ij}$ and $F$ is a $k\times k$ diagonal matrix with diagonal elements $f_{ii}$ such that 
\begin{align}\label{eq: B_and_F}
	b_{ij} = \left\{\begin{array}{l}
	0\; \mbox{ if }\; j \notin N(i)\;; \\
	\frac{\rho}{1 + (n_{<i}-1)\rho^2}\; \mbox{ if }\; i=2,3,\ldots,k\;,\; j\in N(i)\;	
\end{array}	 \right. \quad \mbox{and}\quad f_{ii} = \frac{1 + (n_{<i}-1)\rho^2}{1-\rho^2}\; i=1,2\ldots,k\;,
\end{align}
where $n_{<i}$ is the number of members in $N(i)$. The above definition of $b_{ij}$ is consistent with the lower-triangular structure of $B$ because $j\notin N(i)$ for any $j\geq i$. The derivation of $B$ and $F$ as functions of a spatial correlation parameter $\rho$ is based upon forming local autoregressive models on embedded spanning trees of subgraphs of $\calG$ \citep{datta2018spatial}.    

\subsection{A bivariate DAGAR (BDAGAR) model}\label{sec: bivariate_DAGAR}
We now extend the DAGAR to the bivariate case, where we jointly model two cancers across regions. Let $w_i = (w_{i1}, w_{i2}, \dots, w_{ik})^\top$ be the spatial random effect vector for disease $i$, where $w_{ij}$ refers to the spatial random effect for disease $i$ in region $j$. We will build a hierarchical model,
\begin{equation}\label{eq: bivariate_DAGAR}
p(w_1,w_2) = N(w_1\given 0, \tau_1 Q_1(\rho_1)) \times N(w_2\given A_{21}w_1, \tau_{2} Q_{2}(\rho_2))\;, 
\end{equation}
where $N(\cdot\given \mu, Q)$ denotes a normal density with mean $\mu$ and precision matrix $Q$. The precision matrices $\tau_i Q_i(\rho_i)$ for $i=1,2$ are the DAGAR precision matrices formed with the entries of $B$ and $F$ described in \eqref{eq: B_and_F} with $\rho_i$. Therefore, in \eqref{eq: bivariate_DAGAR} we model $w_1$ as a univariate DAGAR and $w_2$ conditional on $w_1$ also as a DAGAR. Each disease has its own distribution and there are two spatial autocorrelation parameters ($\rho_1$ and $\rho_2$) corresponding to the two diseases. This ensures that spatial associations specific to each disease will be captured. 

The matrix $A_{21}$ models the association between the two diseases. We use a parametric form $A_{21} = \eta_0 I_k + \eta_1 M$, where $M$ is the binary adjacency matrix of the geographic map, i.e., $m_{ij}=1$ if $i\sim j$ and $0$ otherwise. The joint distribution of $w = (w_1^{\top}, w_2^{\top})^{\top}$ is now derived from \eqref{eq: bivariate_DAGAR} as $w \sim N(0, Q_w)$, where the precision matrix $Q_w$ is
\begin{equation}\label{eq: bivariate_DAGAR_precision}
Q_w = \begin{bmatrix} 
\tau_1 Q_1(\rho_1) + \tau_2 A_{21}^{\top}Q_2(\rho_2)A_{21} & \tau_2 A_{21}^{\top} Q_2(\rho_2) \\ 
\tau_2 Q_2(\rho_2)A_{21} & \tau_2 Q_2(\rho_2) \end{bmatrix}                                
\end{equation}
and the covariance matrix $Q_w^{-1}$ is
\begin{equation}\label{eq: bivariate_DAGAR_covariance}
 Q_w^{-1} = \begin{bmatrix} \tau_1^{-1}Q_1^{-1}(\rho_1) & \tau_1^{-1} Q_{1}^{-1}(\rho_1)A_{21}^{\top} \\\tau_1^{-1} A_{21} Q_1^{-1}(\rho_1) & \tau_1^{-1} A_{21} Q_1^{-1}(\rho_1)A_{21}^{\top} + \tau_2^{-1} Q_{2}^{-1}(\rho_2) \end{bmatrix}\;.                                           
\end{equation}
We call a normal distribution with the above precision, or covariance, matrix, the BDAGAR model. The interpretation of $\rho_1$ and $\rho_2$ is clear: $\rho_1$ measures the spatial association for the first cancer, while $\rho_2$ is the residual spatial correlation in the second cancer after accounting for the first cancer. Similarly, $\tau_1$ is the spatial precision parameter for the first cancer, while $\tau_2$ is the residual precision for the second cancer after accounting for the first.

\subsection{Model Implementation} \label{Implementation}
\noindent Let $y_{ij}$ be our outcome of interest corresponding to cancer $i$ in region $j$. We will assume that $y_{ij}$ is a continuous variable, e.g., incidence rates, that is related to a set of explanatory variables through the regression model, 
\begin{equation}\label{eq: spatial_regression}
y_{ij} = x_{ij}^\top\beta_i + w_{ij} + \epsilon_{ij}\;,
\end{equation}
where $x_{ij}$ is a $p_i\times 1$ vector of explanatory variables specific to cancer $i$ within region $j$, $\beta_i$ is the slopes corresponding to cancer $i$, $w_{ij}$'s are the spatial effects that collectively follow the bivariate DAGAR distribution described in Section~\ref{sec: bivariate_DAGAR}, and $\epsilon_{ij} \stackrel{ind}{\sim} N(0,1/\sigma_i^2)$ capture additional heterogeneity and variability independent of spatial variation, where $\sigma_i^2$ is the residual variance for cancer $i$. The regression model is extended to the following specific Bayesian hierarchical framework with the posterior distribution $p(\beta, w, \eta, \rho, \tau, \sigma \given y)$ proportional to
\begin{align}\label{eq: bivariate_DAGAR_bhm}
& p(\rho) \times p(\eta) \times \prod_{i=1}^2 \left\{IG(1/\tau_i\given a_{\tau_i}, b_{\tau_i}) \times IG(\sigma_i^2 \given a_{\sigma}, b_{\sigma}) \times N(\beta_i\given \mu_{\beta_i}, V_{\beta_i}^{-1})\right\} \nonumber\\ 
&\qquad \times N(w\given 0, Q_{w}) \times \prod_{i=1}^2\prod_{j=1}^k N(y_{ij}\given x_{ij}^{\top}\beta_i + w_{ij},\sigma_i^2)\;,
\end{align} 
where $\beta = \{\beta_1, \beta_2\}$, $\tau=\{\tau_1,\tau_2\}$, $\sigma = \{\sigma^2_1, \sigma^2_2\}$ and $\eta = \{\eta_0,\eta_1\}$, and $IG(\cdot\given a,b)$ is the inverse-gamma distribution with shape and rate parameters $a$ and $b$, respectively.

We sample the parameters from the posterior distribution in \eqref{eq: bivariate_DAGAR_bhm} using Markov chain Monte Carlo (MCMC) with Gibbs sampling and random walk metropolis \citep{gamerman2006markov} as implemented in the \texttt{rjags} package within the \texttt{R} statistical computing environment. To compare and assess models, we use the Widely Applicable Information Criterion (WAIC) \citep{watanabe2010asymptotic,gelman2014understanding}, which is computed as
\begin{align*}
\mbox{WAIC} = -2\widehat{elppd} =  -2(\widehat{lppd} - \hat{p}_{WAIC})\;,
\end{align*}
where $\widehat{elppd}$ is the expected log point-wise predictive density for a new dataset and $\hat{p}_{WAIC}$ is the estimated effective number of parameters, which is sum of posterior variance of the log predictive density for each data point. WAIC is easy to compute using posterior samples.

\section{Results}\label{sec: Data_Analysis}
We analyze a data set extracted from the SEER$^*$Stat database using the SEER$^*$Stat statistical software \citep{seer}. We consider 2 cancers, lung and esophagus, where the outcome is the crude incidence rates per 100,000 population in the years from 2012 to 2016 across 58 counties in California, USA. County-level explanatory variables for each cancer are available in the same years and include percentages of residents younger than 18 years old (young$_{ij}$), older than 65 years old (old$_{ij}$), with education level below high school (edu$_{ij}$) , percentages of unemployed residents (unemp$_{ij}$), black residents (black$_{ij}$), male residents (male$_{ij}$), uninsured residents (uninsure$_{ij}$), and percentages of families below the poverty threshold (poverty$_{ij}$).

\begin{figure}[H]
	\centering
	\includegraphics[width=173mm]{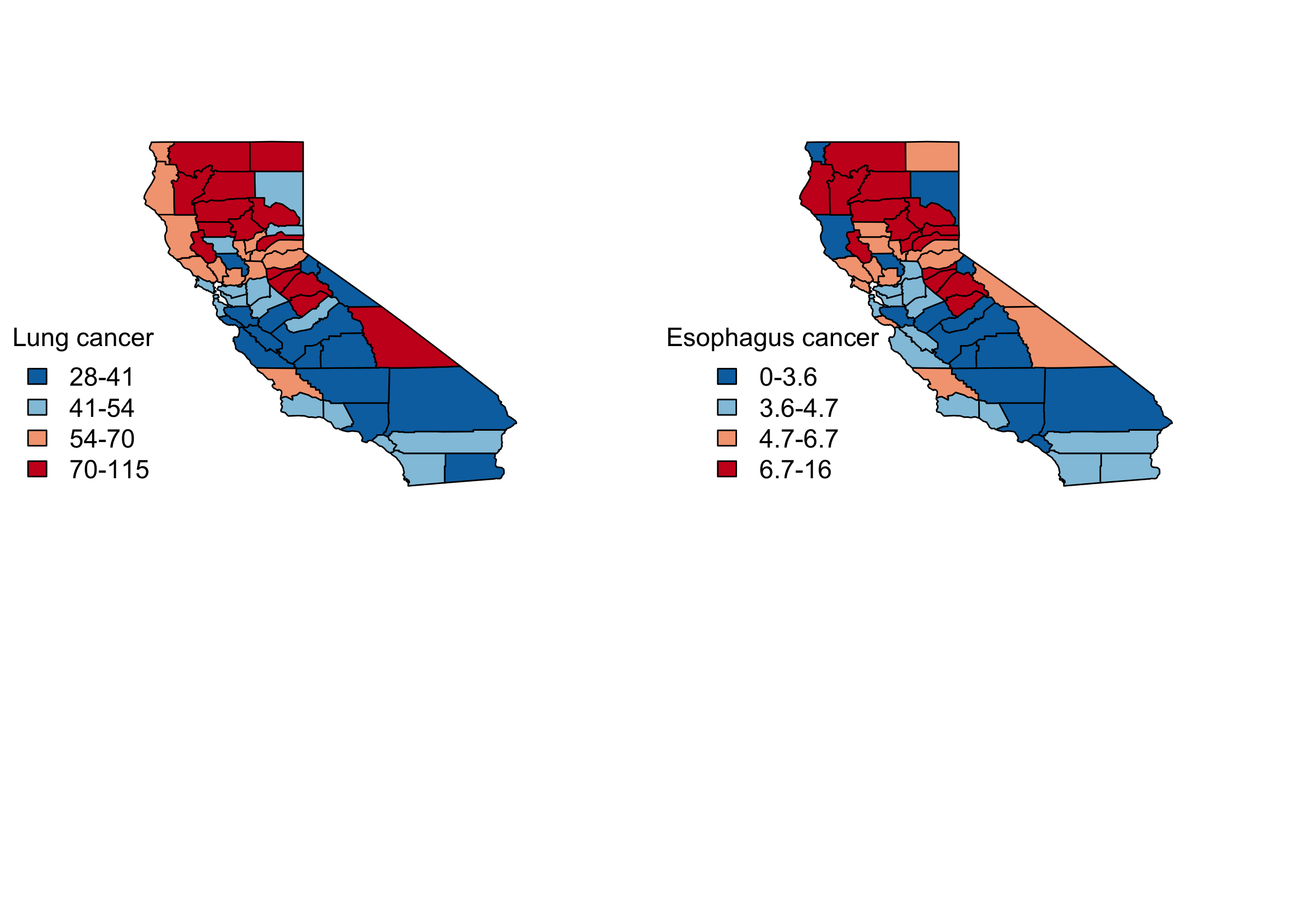}
	\caption{Maps of incidence rates per 100,000 population for lung and esophagus cancer in California.}\label{fig: cancer_incidence_maps}
\end{figure}

We analyzed this data set using the Bayesian hierarchical model \eqref{eq: bivariate_DAGAR_bhm}. The county-level maps of the raw incidence rates per 100,000 population for the two cancers are shown in Figure~\ref{fig: cancer_incidence_maps}. The maps exhibit the evidence of correlation across space and between cancers. Cutoffs for the different levels of incidence rates are quantiles for each cancer. For both lung and esophagus cancer, in general, incidence rates are higher in counties located in the northern areas than those in southern part. The four counties in the center including Amador, Calaveras, Tuolumne and Mariposa have relatively high incidence rates compared to the neighboring counties. Overall, counties with similar levels of incidence rates tend to depict some spatial clustering. 

For our analysis, we specified the following prior distribution,
\begin{align}\label{eq: priors}
p(\eta,\rho,\tau,\sigma,w) &= \prod_{i=1}^2 Unif(\rho_i\given 0,1)\times \prod_{i=0}^1N(\eta_i\given 0, 10^2)\times \prod_{i=1}^2 N(\beta\given 0, 10^3)\nonumber\\ 
&\times \prod_{i=1}^2IG(1/\tau_i\given 2,0.1)\times \prod_{i=1}^2 IG(\sigma_i^2\given 2,1) \times N(w\given 0, Q_w(\tau,\rho))\;,   
\end{align} 
where $Unif(\cdot\given a,b)$ denotes the Uniform density over $(0,1)$ and $Q_w(\tau,\rho)$ is the BDAGAR precision matrix of $w$ given in [\ref{eq: bivariate_DAGAR_precision}]. 

\begin{figure}[H]
	\centering
	\includegraphics[width=173mm]{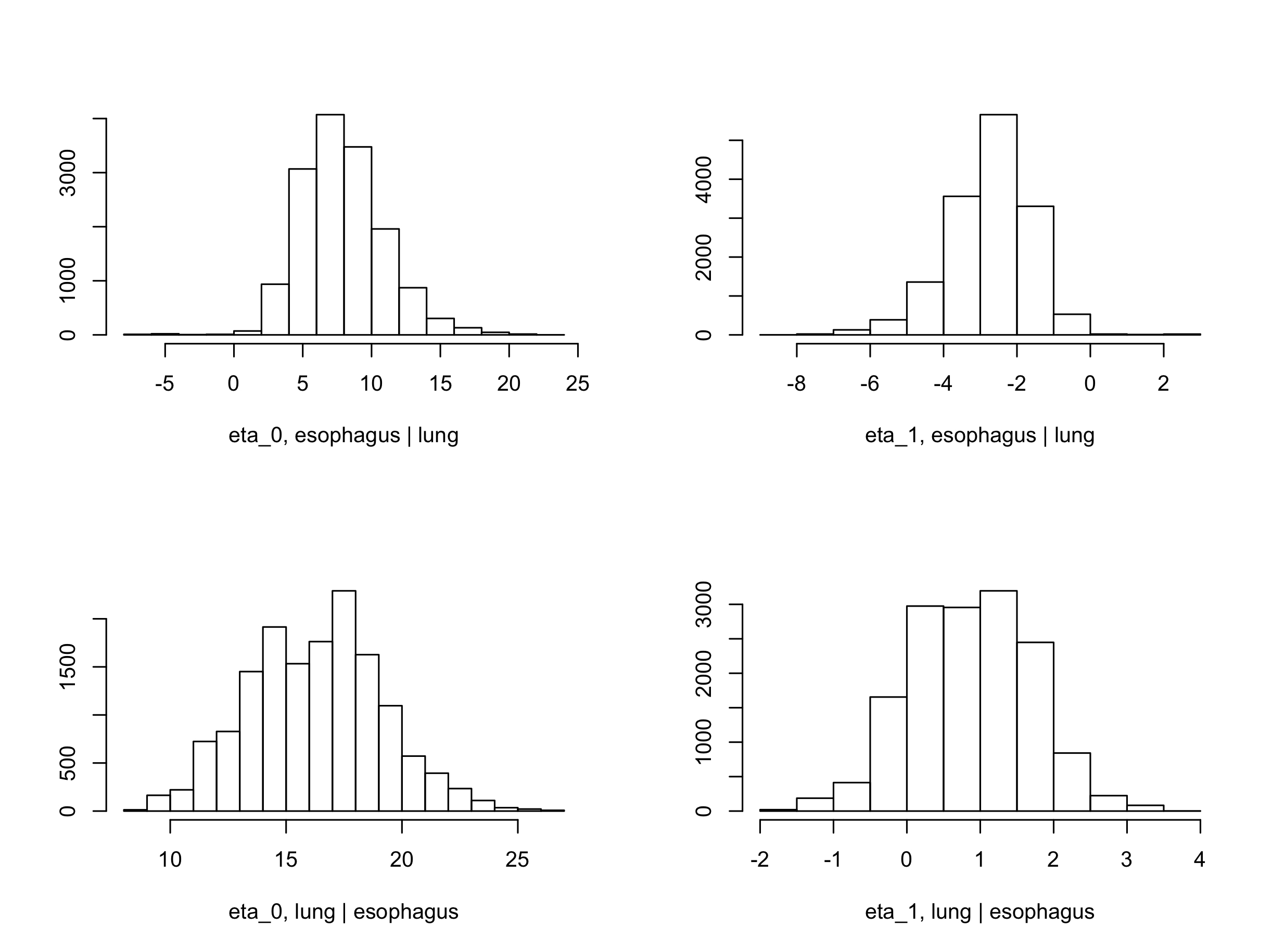}
	\caption{Posterior samples of linking parameters $\eta_{0}, \eta_{1}$ from BDAGAR model.}\label{fig: Posterior_distribution}
\end{figure}

We fit the BDAGAR model using the two different cancer orders, i.e. $[\mbox{esophagus}] \times [\mbox{lung}\given\mbox{esophagus}]$ and the reverse ordering $[\mbox{lung}] \times [\mbox{esophagus}\given\mbox{lung}]$. We will refer to these orderings simply as $[\mbox{lung}\given\mbox{esophagus}]$ and $[\mbox{esophagus}\given\mbox{lung}]$, respectively. Table~\ref{tab: model_comparison} presents measures for model fit using the WAIC. We also compare BDAGAR with the ``Generalized Multivariate Conditional Autoregression (GMCAR)'' models \citep{jin2005generalized}. In both BDAGAR and GMCAR models, the conditional order $[\mbox{esophagus}] \times [\mbox{lung}\given\mbox{esophagus}]$ has a smaller WAIC (hence better fit to the data) than the reverse ordering. Meanwhile, within each order, BDAGAR seems to excel over the GMCAR with lower scores in both model fit and effective number of parameters, as seen in the values of $\widehat{elppd}$ and $\hat{p}_{WAIC}$, respectively. The preference of WAIC for $[\mbox{lung}\given\mbox{esophagus}]$ is also corroborated by the posterior distribution of $\eta_{0}$ and $\eta_{1}$ from BDAGAR shown in Figure~\ref{fig: Posterior_distribution}. In $[\mbox{esophagus}\given\mbox{lung}]$, the parameter $\eta_{1}$ has posterior median of $-2.73$ and a $95\%$ credible interval $(-5.27,-0.81)$. This shows significant negative values that offset part of the significant positive effect of $\eta_{0}$ with a median of $7.92$ and a $95\%$ credible interval of $(3.15,14.60)$. For $[\mbox{lung}\given\mbox{esophagus}]$, $\eta_{0}$ is significantly positive with a median of $16.27$ and $95\%$ credible interval of $(10.95,22.09)$, while $\eta_{1}$ tends to be positive with a median of $0.87$ but with a $95\%$ credible interval $(-0.69,2.43)$ that includes $0$. Consequently, we present the following results and analysis for $[\mbox{lung}\given\mbox{esophagus}]$ which seems to be the preferred model.

Table~\ref{tab: posterior_estimates} summarizes the parameter estimates from the BDAGAR model corresponding to $[\mbox{lung}\given\mbox{esophagus}]$. For fixed effects, the increasing percentage of residents younger than 18 years old significantly reduces the incidence rate for esophagus cancer, while the percentage of residents older than 65 years old has a significantly opposite effect for both esophagus and lung cancer. The increase in the percentage of unemployment also augments the incidence rate of lung cancer significantly. Turning to spatial correlations, $\rho_1$ measures the residual spatial correlation (posterior mean $0.11$) for esophagus cancer after accounting for the explanatory variables and $\rho_2$ measures the spatial correlation (posterior mean $0.47$) for lung cancer after accounting for the explanatory variables and also the effect of esophagus cancer. The small point estimates and narrower credible interval for $\rho_1$ indicate greater confidence in weaker spatial correlation for esophagus cancer; the moderate value of $\rho_2$ and a wider credible interval suggest higher spatial correlation for lung cancer. Turning to the spatial precision of random effects for each cancer, the estimates of $\{\tau_1,\tau_2\}$ are indicative of esophagus cancer having larger variability, although we must keep in mind that $\tau_2$ is the conditional marginal precision for lung cancer after accounting for esophagus cancer and, therefore, may not be directly comparable to $\tau_1$.

\begin{figure}[H]
	\centering
	\includegraphics[width=173mm]{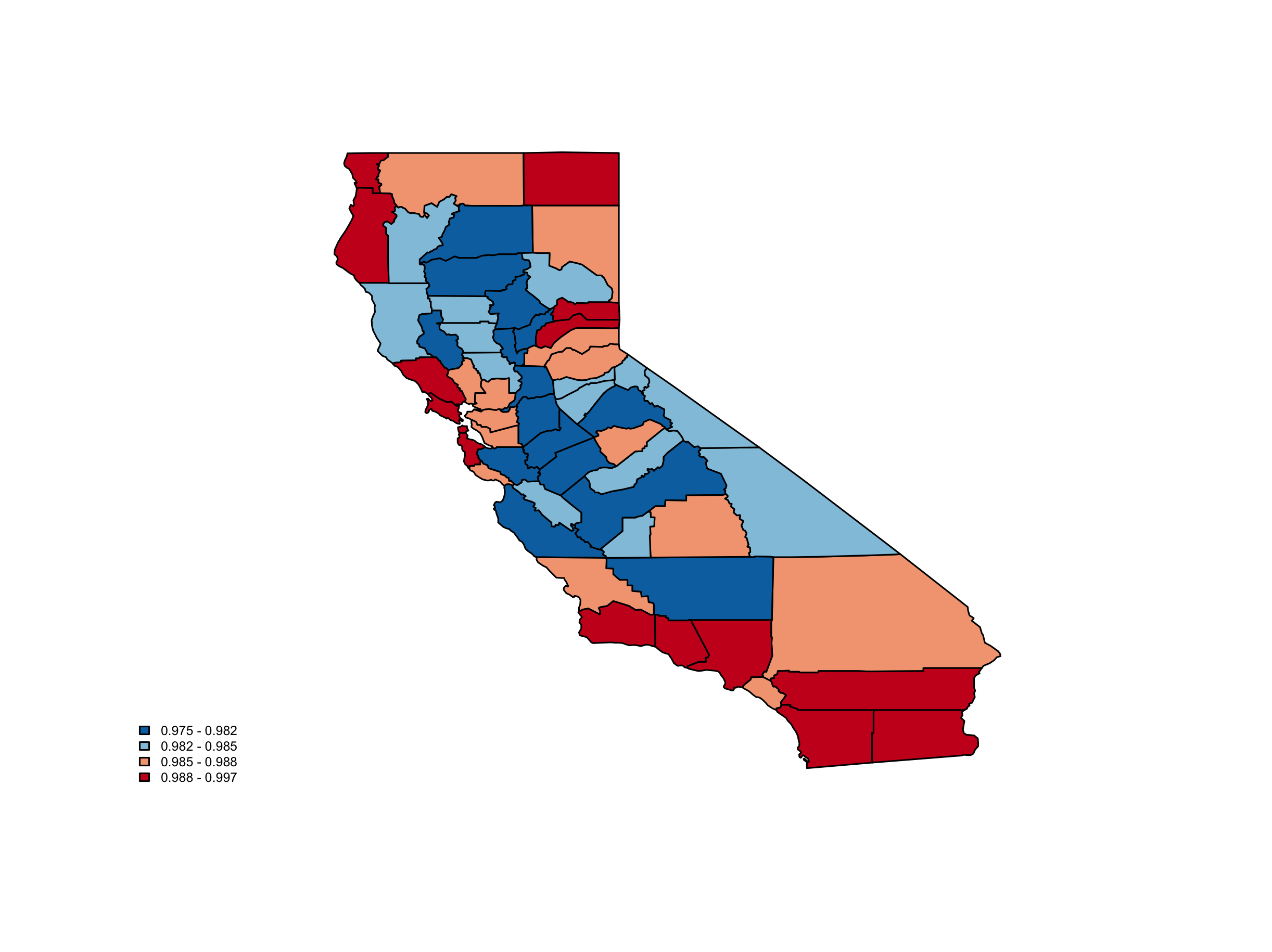}
	\caption{Estimated correlation between lung and esophagus cancer in each of 58 counties of California.}\label{fig: posterior_correlations}
\end{figure}

Figure~\ref{fig: posterior_correlations} shows the estimated correlation between lung and esophagus cancer in each of 58 counties. This map also seems to be consistent with the estimates of $\eta$. Correlations between lung and esophagus cancers in all counties are significantly positive with large means at around $0.97 - 1$ which are due to the highly positive values in $\eta_{0}$. This indicates that esophagus cancer is highly correlated with lung cancer. However, in general, the correlation between the two cancers increases slightly from the center to marginal areas, especially for those with fewer counties in the neighborhood. 

\begin{figure}[H]
	\centering
	\includegraphics[width=173mm]{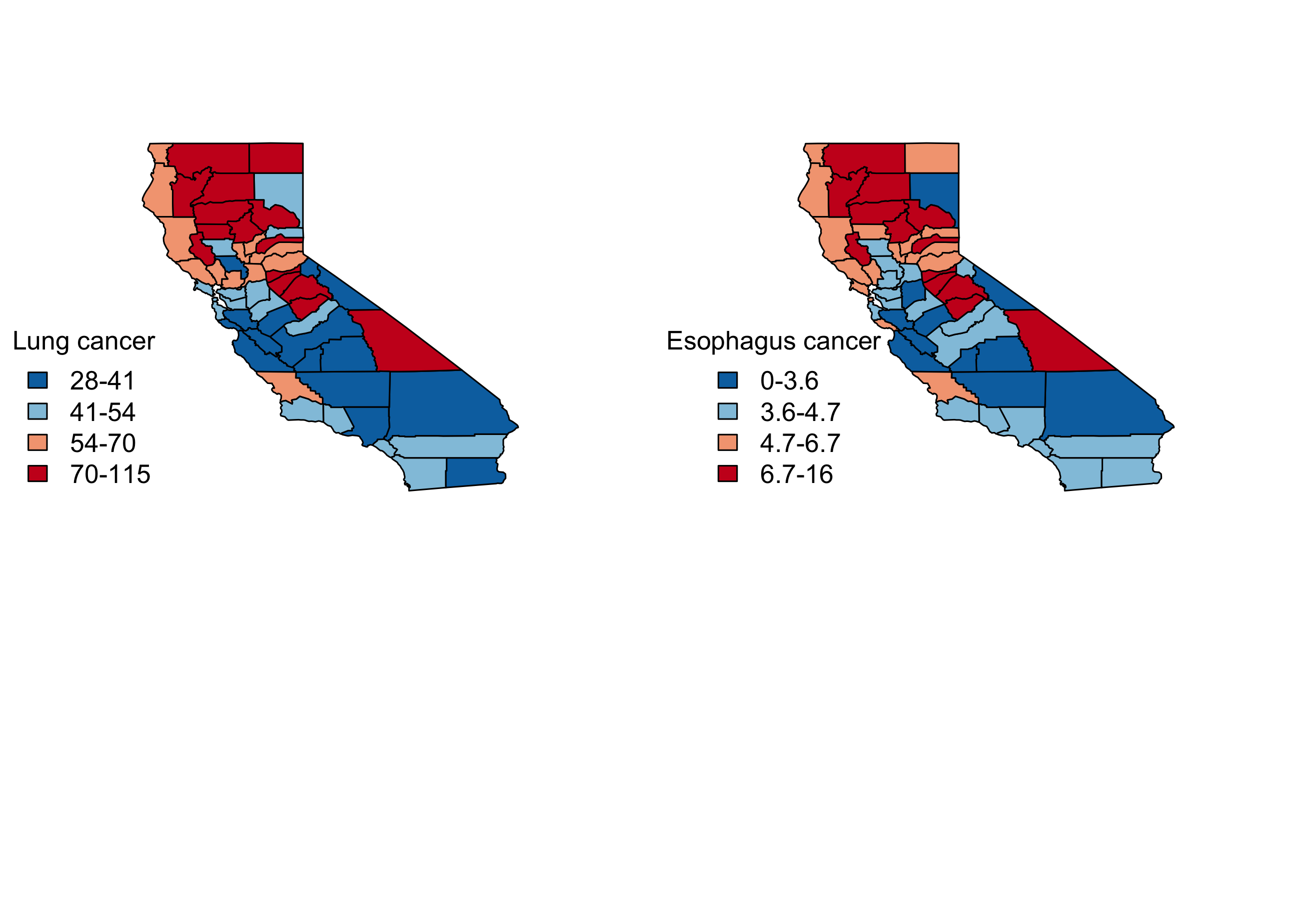}
	\caption{Maps of posterior mean incidence rates per 100,000 population for lung and esophagus cancer in California}\label{fig: posterior_incidence}
\end{figure}

Finally, Figure~\ref{fig: posterior_incidence} provides further visual corroboration of the goodness of fit for the BDAGAR mode corresponding to $[\mbox{lung}\given\mbox{esophagus}]$. Here, we see that the posterior mean of the incidence rates for lung and esophagus cancer are very consistent with the raw incidence rates shown in Figure~\ref{fig: cancer_incidence_maps}. 

\begin{table}[H]
	\centering
	\caption{Model comparison using WAIC statistics for cancer data analysis}\label{tab: model_comparison}
	\begin{tabular}{cccc}
		\hline
		Model & \multicolumn{1}{l}{lppd} & \multicolumn{1}{l}{$p_{WAIC}$} & \multicolumn{1}{l}{WAIC} \\
		\hline
		BDAGAR (esophagus $|$ lung) & -273.87  & 44.62 & 636.99 \\
		BDAGAR (lung $|$ esophagus) & -158.25 & 50.27 & 417.05 \\
		GMCAR (esophagus $|$ lung)& -282.76 & 48.23 & 661.97\\
		GMCAR (lung $|$ esophagus)& -158.76 & 52.33 & 422.18\\
		\hline
	\end{tabular}%
\end{table}%

\begin{table}[H]
	\centering
	\caption{Parameter estimates (posterior means) for the California cancer incidence rate data from BDAGAR model. The numbers inside braces indicates the lower and upper bounds for the 95$\%$ credible intervals.} \label{tab: posterior_estimates}
	\begin{tabular}{ccccc}
		\hline
		Parameters & Esophagus cancer & Lung cancer\\
		\hline
		intercept & 15.40 (0.45, 30.36) & 0.71 (-52.61, 55.17) \\
		young &  -0.24 (-0.48, -0.01) & -0.93 (-2.16, 0.31)  \\
		old   & 0.23 (0.06, 0.41) & 2.93 (1.91, 3.98) \\
		edu   & -0.02 (-0.15, 0.10) & -0.39 (-1.22, 0.44) \\
		unemp & 0.12 (-0.05, 0.29) & 1.41 (0.22, 2.61)\\
		black & 0.16 (-0.09, 0.41) & 0.87 (-0.84, 2.54)\\
		male  & -0.06 (-0.23, 0.10) & 0.01 (-1.15, 1.16)\\
		uninsure &-0.22 (-0.45, 0.02) & 0.24 (-0.78, 1.24)\\
		poverty & 0.38 (-0.31, 1.07) & 0.47 (-3.98, 5.01) \\
		$\sigma_{ei}^2$ & 2.31 (1.55, 3.40) & 0.88 (0.17, 3.54) \\
		$\tau_i^2$ & 2.00 (0.76, 3.84) & 19.72 (2.25, 56.18)\\
		$\rho_i$ & 0.11 (0.00, 0.31) & 0.47 (0.02, 0.97)\\
		\hline
	\end{tabular}%
	\label{tab:addlabel}%
\end{table}%

\section{Discussion}\label{sec: Discussion}
We have extended a recently proposed class of DAGAR models \citep{datta2018spatial} for univariate disease mapping to bivariate ``BDAGAR'' models that can be applied to estimate spatial correlations for two correlated cancers. The BDAGAR model retains the interpretation of DAGAR models clearly separating the spatial correlation for each cancer from any inherent or endemic association between the two cancers. The BDAGAR model can still be efficiently computed using MCMC algorithms. Our analysis of incidence rates from lung and esophagus cancer demonstrates the efficiency of BDAGAR and its improved performance, as measured by WAIC, over existing alternatives such as the GMCAR models.

While we have restricted our attention only to cancer incidence rates, BDAGAR models can also be used with time-to-event data to investigate geographical patterns in the hazard function. For example, each patient in a study may provide multiple survival times from the onset of each of two cancers along with his or her county of residence. The BDAGAR model can become an excellent alternative to CAR and different MCAR models in spatial survival analysis \citep{banerjee2003frailty,carlin2003hierarchical,banerjee2005semiparametric, cooner2006modelling,  diva2008parametric}.

Finally, the BDAGAR models developed here proceeds from conditional specifications. Concerns may arise over the ordering of the variables in the hierarchical approach. While in the case of a few cancers, such as 2 in our case, one can evaluate models arising from the different orders, this strategy will become cumbersome with several cancers. For instance, even with $4$ cancers, we will have $24$ different models that will need to be evaluated and compared. This becomes impractical. A joint modeling approach, analogous to order-free MCAR models as in \citep{jin2007order}, can build rich spatial structures from linear transformations of simpler latent variables. For instance, we can develop alternate multivariate DAGAR, or MDAGAR models, using $w = \Lambda f$, where $\Lambda$ is a suitably specified square matrix and $f$ is a latent vector whose components follow independent univariate DAGAR distributions. Note that by modeling the joint distribution, the incompatibility of conditional model building (i.e., different joint distributions for different orderings) is avoided. However, the issue of the
identifiability of $\Lambda$ is raised, and careful specification of its structure is needed. These approaches will be further investigated elsewhere.

\section*{Acknowledgments}
The work of the first and second authors have been supported in part by the Division of Mathematical Sciences (DMS) of the National Science Foundation (NSF) under grant 1916349 and by the National Institute of Environmental Health Sciences (NIEHS) under grants R01ES030210 and 5R01ES027027. 

%\newpage

\bibliographystyle{vancouver}
%\bibliography{Banerjee}

\begin{thebibliography}{10}
	
	\bibitem{koch2005cartographies}
	Koch T.
	\newblock Cartographies of disease: maps, mapping, and medicine.
	\newblock Esri Press Redlands, CA; 2005.
	
	\bibitem{schaible2013indirect}
	Schaible WL.
	\newblock Indirect estimators in US federal programs. vol. 108.
	\newblock Springer Science \& Business Media; 2013.
	
	\bibitem{banerjee2014hierarchical}
	Banerjee S, Carlin BP, Gelfand AE.
	\newblock Hierarchical modeling and analysis for spatial data.
	\newblock CRC Press, Boca Raton, FL; 2014.
	
	\bibitem{best2005comparison}
	Best N, Richardson S, Thomson A.
	\newblock A comparison of Bayesian spatial models for disease mapping.
	\newblock Statistical Methods in Medical Research. 2005;14(1):35--59.
	
	\bibitem{waller2010handbook}
	Waller L, Carlin B. Handbook Of Spatial Statistics. P. Diggle, M. Fuentes, AE
	Gelfand, and P. Guttorp, Boca Raton, FL: Taylor and Franci.
	\newblock USA: Chapman and Hall CRC Press; 2010.
	
	\bibitem{waller2004applied}
	Waller LA, Gotway CA.
	\newblock Applied spatial statistics for public health data. vol. 368.
	\newblock John Wiley \& Sons; 2004.
	
	\bibitem{lawson2013statistical}
	Lawson AB.
	\newblock Statistical methods in spatial epidemiology.
	\newblock John Wiley \& Sons; 2013.
	
	\bibitem{rueheld04}
	Rua H, Held L.
	\newblock Gaussian Markov Random Fields : Theory and Applications.
	\newblock Monographs on statistics and applied probability. Chapman and
	Hall/CRC Press, Boca Raton, FL; 2005.
	\newblock Available from: \url{http://opac.inria.fr/record=b1119989}.
	
	\bibitem{besag1974spatial}
	Besag J.
	\newblock Spatial interaction and the statistical analysis of lattice systems.
	\newblock Journal of the Royal Statistical Society: Series B (Methodological).
	1974;36(2):192--225.
	
	\bibitem{besag1991bayesian}
	Besag J, York J, Molli{\'e} A.
	\newblock Bayesian image restoration, with two applications in spatial
	statistics.
	\newblock Annals of the Institute of Statistical Mathematics. 1991;43(1):1--20.
	
	\bibitem{anselin1988lagrange}
	Anselin L.
	\newblock Lagrange multiplier test diagnostics for spatial dependence and
	spatial heterogeneity.
	\newblock Geographical Analysis. 1988;20(1):1--17.
	
	\bibitem{kissling2008spatial}
	Kissling WD, Carl G.
	\newblock Spatial autocorrelation and the selection of simultaneous
	autoregressive models.
	\newblock Global Ecology and Biogeography. 2008;17(1):59--71.
	
	\bibitem{datta2018spatial}
	Datta A, Banerjee S, Hodges JS, et~al.
	\newblock Spatial disease mapping using directed acyclic graph auto-regressive
	(DAGAR) models.
	\newblock Bayesian Analysis. 2018;.
	
	\bibitem{knorr2001shared}
	Knorr-Held L, Best NG.
	\newblock A shared component model for detecting joint and selective clustering
	of two diseases.
	\newblock Journal of the Royal Statistical Society: Series A (Statistics in
	Society). 2001;164(1):73--85.
	
	\bibitem{kim2001bivariate}
	Kim H, Sun D, Tsutakawa RK.
	\newblock A bivariate Bayes method for improving the estimates of mortality
	rates with a twofold conditional autoregressive model.
	\newblock Journal of the American Statistical Association.
	2001;96(456):1506--1521.
	
	\bibitem{gelfand2003proper}
	Gelfand AE, Vounatsou P.
	\newblock Proper multivariate conditional autoregressive models for spatial
	data analysis.
	\newblock Biostatistics. 2003;4(1):11--15.
	
	\bibitem{carlin2003hierarchical}
	Carlin BP, Banerjee S.
	\newblock Hierarchical multivariate CAR models for spatio-temporally correlated
	survival data.
	\newblock Bayesian Statistics. 2003;7(7):45--63.
	
	\bibitem{held2005towards}
	Held L, Nat{\'a}rio I, Fenton SE, et~al.
	\newblock Towards joint disease mapping.
	\newblock Statistical Methods in Medical Research. 2005;14(1):61--82.
	
	\bibitem{jin2005generalized}
	Jin X, Carlin BP, Banerjee S.
	\newblock Generalized hierarchical multivariate CAR models for areal data.
	\newblock Biometrics. 2005;61(4):950--961.
	
	\bibitem{jin2007order}
	Jin X, Banerjee S, Carlin BP.
	\newblock Order-free co-regionalized areal data models with application to
	multiple-disease mapping.
	\newblock Journal of the Royal Statistical Society: Series B (Statistical
	Methodology). 2007;69(5):817--838.
	
	\bibitem{diva2008parametric}
	Diva U, Dey DK, Banerjee S.
	\newblock Parametric models for spatially correlated survival data for
	individuals with multiple cancers.
	\newblock Statistics in Medicine. 2008;27(12):2127--2144.
	
	\bibitem{zhang2009smoothed}
	Zhang Y, Hodges JS, Banerjee S.
	\newblock Smoothed ANOVA with spatial effects as a competitor to MCAR in
	multivariate spatial smoothing.
	\newblock The Annals of Applied Statistics. 2009;3(4):1805.
	
	\bibitem{martinez2013general}
	Martinez-Beneito MA.
	\newblock A general modelling framework for multivariate disease mapping.
	\newblock Biometrika. 2013;100(3):539--553.
	
	\bibitem{mari2014smoothed}
	Mar{\'\i}-DellâOlmo M, Martinez-Beneito MA, Gotsens M, et~al.
	\newblock A smoothed ANOVA model for multivariate ecological regression.
	\newblock Stochastic Environmental Research and Risk Assessment.
	2014;28(3):695--706.
	
	\bibitem{gamerman2006markov}
	Gamerman D, Lopes HF.
	\newblock Markov chain Monte Carlo: stochastic simulation for Bayesian
	inference.
	\newblock Chapman and Hall/CRC; 2006.
	
	\bibitem{watanabe2010asymptotic}
	Watanabe S.
	\newblock Asymptotic equivalence of Bayes cross validation and widely
	applicable information criterion in singular learning theory.
	\newblock Journal of Machine Learning Research. 2010;11(Dec):3571--3594.
	
	\bibitem{gelman2014understanding}
	Gelman A, Hwang J, Vehtari A.
	\newblock Understanding predictive information criteria for Bayesian models.
	\newblock Statistics and Computing. 2014;24(6):997--1016.
	
	\bibitem{seer}
	Step 1: Calculating Age-adjusted Rates - SEER*Stat Tutorials;.
	\newblock Available from:
	\url{https://seer.cancer.gov/seerstat/tutorials/aarates/step1.html}.
	
	\bibitem{banerjee2003frailty}
	Banerjee S, Wall MM, Carlin BP.
	\newblock Frailty modeling for spatially correlated survival data, with
	application to infant mortality in Minnesota.
	\newblock Biostatistics. 2003;4(1):123--142.
	
	\bibitem{banerjee2005semiparametric}
	Banerjee S, Dey DK.
	\newblock Semiparametric proportional odds models for spatially correlated
	survival data.
	\newblock Lifetime Data Analysis. 2005;11(2):175--191.
	
	\bibitem{cooner2006modelling}
	Cooner F, Banerjee S, McBean AM.
	\newblock Modelling geographically referenced survival data with a cure
	fraction.
	\newblock Statistical Methods in Medical Research. 2006;15(4):307--324.
	
\end{thebibliography}

\end{document}